# Internet of Things Platform Service Supply Innovation: Exploring the Impact of Overconfidence

Xiufeng Li[a,*], Zefang Li[b]

[a] School of Business, Nanjing Normal University, Nanjing 210023, China;
[b] School of Management Engineering, Qingdao University of Technology, Shandong, Qingdao, 266520, China

**Abstract**: This paper explores the impact of manufacturers' overconfidence on their collaborative innovation with platforms in the Internet of Things (IoT) environment by constructing a game model. It is found that in both usage-based and revenue-sharing contracts, manufacturers' and platforms' innovation inputs, profit levels, and pricing strategies are significantly affected by the proportion of non-privacy-sensitive customers, and grow in tandem with the rise of this proportion. In usage-based contracts, moderate overconfidence incentivizes manufacturers to increase hardware innovation investment and improve overall supply chain revenues, but may cause platforms to reduce software innovation; under revenue-sharing contracts, overconfidence positively incentivizes hardware innovation and pricing more strongly, while platform software innovation varies nonlinearly depending on the share ratio. Comparing the differences in manufacturers' decisions with and without overconfidence suggests that moderate overconfidence can lead to supply chain Pareto improvements under a given contract. This paper provides new perspectives for understanding the complex interactions between manufacturers and platforms in IoT supply chains, as well as theoretical support and practical guidance for actual business decisions.

# 1.Introduction

The development of Internet of Things (IoT) technology has pushed smart products to gradually become mainstream consumer products, which bring consumers a more convenient and intelligent life experience(Ladeira et al., 2025), while consumers are increasingly concerned about the privacy of IoT technology in smart products(Jaspers and Pearson, 2022). For example, Amazon Ring was fined $5.6 million by the FTC in 2024 due to internal control failures that caused employees to access users' private videos at will[1]. Thus IoT platforms that leak data due to improper innovation may reduce customer trust and trigger financial losses.

On the other hand, manufacturers and platforms need to invest in R&D to improve product innovation, but differences in cooperation models and benefit sharing between the two parties may lead to inefficiencies in cooperation. The challenge of IoT technology application involves both technology and business model innovation (Ding et al., 2023), and the benefits of cooperation between the two parties are significant, such as IoT platforms can help manufacturers innovate product service systems(Shin et al., 2022), which is more conducive to the realization of seamless integration of hardware and software to enhance the overall level of innovation. For example, IBM spent 200 million U.S. dollars to build the Watson IoT headquarters in Munich, and BMW in-depth cooperation with the development of intelligent driving solutions, BMW team stationed in the borrowing of IBM technology to optimize the vehicle data analysis and other functions.

---

[1] https://www.cbsnews.com/news/amazon-ring-ftc-lawsuit-customer-videos/

In reality, although collaboration can be beneficial to both parties and both parties expect to innovate and maximize their own interests based on the collaboration, the game of profit maximization still needs to be played out in practice. Since innovation often requires high R&D costs, manufacturers often face these costs with the cognitive bias of overconfidence, which can have serious negative consequences. For example, Elon Musk's overconfidence in his acquisition of Twitter, where he failed to conduct adequate due diligence and offered $44 billion at a significantly higher market valuation, led to huge debt and a heavy financial and reputational price for overconfidence[2]; Sigfox's overconfidence in its ultra-narrowband technology led to bets that it would become the core global IoT standard, and to an aggressive expansion of its global network. However, the technology failed to meet the challenges of a diversified market, coupled with high infrastructure costs that were unsustainable, and was eventually declared bankrupt in 2022[3]; Tesla, on the other hand, was overconfident, blurring the boundaries between "Assisted Driving" and "Fully Autonomous Driving" for a long period of time, and exaggerating its technological capabilities, causing users to misjudge its performance. which led to misjudgment by users and resulted in a number of fatal accidents. One of those accidents in Florida in 2025 resulted in a $243 million judgment against it[4]. To explore the impact of manufacturers' overconfidence on themselves and their platforms, this study incorporates manufacturers' cognitive biases while considering platforms' co-innovation with manufacturers. Overconfidence may allow manufacturers to overestimate market demand, over-recognize their own innovation capabilities, and underestimate the risks of collaboration, leading to more aggressive innovation strategies (Wu et al., 2025). However, such decisions may also bring risks, such as wasting resources by investing more in innovation than is actually needed, or ignoring platform opinions to affect the efficiency of cooperation and relationship stability. However, overconfident manufacturers may also have positive effects: first, they are more willing to take innovation risks and invest large amounts of resources in research and development to push forward technological breakthroughs and product innovations, such as Tesla's innovations in the fields of electric cars and self-driving, which are partly due to Musk's lofty goals and bold decision-making; second, they may increase market acceptance of innovative products and accelerate the commercialization of innovative products through active marketing and brand building; third, they may be more willing to cooperate with platforms to promote innovation and commercialize their products; and third, they may be more willing to cooperate with platforms to promote innovation. Secondly, it may enhance market acceptance of innovative products through active marketing and brand building, and accelerate the commercialization of innovative products; thirdly, it may show stronger leadership and decision-making power in cooperation, and push forward the implementation of cooperation projects quickly.

---

[2] https://www.cnn.com/2022/10/28/tech/elon-musk-twitter-deal-close
[3] https://www.eetimes.com/sigfox-files-for-bankruptcy-protection-to-survive-the-pandemic/
[4] https://www.cnn.com/2025/08/01/business/tesla-autopilot-crash-lawsuit

Given the possible risks and opportunities associated with overconfidence, this study aims to delve into the mechanisms by which manufacturers' overconfidence affects their collaborative innovation with platforms. By constructing theoretical mode, we will analyze how overconfidence affects manufacturers' and platforms' innovation decisions and collaboration outcomes under two contracts respectively. At the same time, we will also explore how to effectively manage and channel this cognitive bias in cooperation to achieve a more efficient and stable partnership, in addition to the following questions.

(1) how will the proportion of non-privacy-sensitive customers affect platforms' and manufacturers' innovation decisions and their profits?

(2) how will overconfident manufacturers affect their own and platforms' innovation decisions under different contractual scenarios?

(3) will overconfident manufacturers actually harm their own and the platform's profits?

To answer these questions, we develop a game model considering a supply chain consisting of overconfident smart product manufacturers and Internet platforms. In this case, the manufacturer develops and innovates new product hardware features based on its own smart product features, and the Internet platform improves the smart product features based on the innovation of software services by taking advantage of its scale effect and integrating them with the hardware facilities generated by the manufacturer, taking into account the impact of the manufacturer's overconfidence. In addition, we also consider the possibility that consumers may be resistant to sharing their data due to the existence of Internet platforms, and that consumers may be able to bring additional value to Internet platforms if they are willing to share their data. To the best of our knowledge, previous literature has not yet fully explored the impact of smart device manufacturers' overconfidence in the context of functional supply chains on their and IoT platforms' respective innovations. Therefore, our paper is a first attempt to introduce the marginal value of data brought by smart device manufacturers' overconfidence and non-privacy sensitive customers in order to facilitate IoT innovation using a typical game model. Overall, this study is of strong academic interest, taking into account the reality that manufacturers may be imperfectly rational and the value of data sharing arising from the application of Internet platform services.

This study provides important theoretical and practical insights into the collaborative innovation decisions between IoT platforms and manufacturers. First, we construct an innovative game-theoretic model that incorporates the manufacturer's overconfident behavioral characteristics and the proportion of non-privacy-sensitive customers into the analytical framework to examine the innovative interaction mechanisms of supply chain members under two contractual models. It is found that a manufacturer's moderate overconfidence can enhance the overall supply chain performance. When a manufacturer exhibits moderate overconfidence, it significantly increases its hardware innovation investment, and this irrational behavior may instead bring unexpected benefits to the whole supply chain. Especially under different cooperation models, this effect presents differentiated characteristics: in a fixed-fee royalty contract, the platform's software innovation will reduce the level of innovation with the manufacturer's overconfidence due to the fixed revenue received, but at the same time, it is also lowering the fixed revenue it receives along with it in order to balance out the reduced level of innovation; whereas, in a revenue-sharing contract, the platform adjusts its innovation strategy according to its own share of benefits, and under a certain percentage, the platform will adjust its innovation strategy according to its own share of benefits. strategy, and at a given rate, the platform's innovation may rise as the manufacturer's overconfidence level rises. Second, this study reveals the key role of the proportion of non-privacy-sensitive customers in regulating innovation inputs and profit sharing, and when the revenue-sharing contract split ratio is in a certain interval, manufacturers and platforms are more favorable in choosing usage-based contracts when there are fewer non-privacy-sensitive customers, which expands the boundaries of existing research on IoT service supply chains. Finally, the findings provide concrete guidance for the cooperation practices between IoT platforms and manufacturers: firms can guide the investment of innovation resources by rationally designing contract terms, and they need to adjust their cooperation strategies according to the characteristics of market customer structure. These findings not only fill the research gap of behavioral operations management in the field of IoT service supply chain, but also provide a decision-making basis for enterprises to formulate technological innovation strategies.

## 2. Literature Review

### 2.1 Internet of Things Innovation

On the issue of business model innovation and corporate value creation, Haaker et al. (2021) explore the application of IoT in business model innovation in Vietnamese firms through a case study, emphasizing the importance of synergistic innovations in technology, management, services, and finance; Markfort et al. (2022), through a mixed-methods study, find that corporate IoT platforms three modes of business model innovation: platform piloting, platform revenue generation, and platform orchestration; Del Giudice(2016) explores the role of IoT as a technological innovation in business process management, emphasizing its potential to foster knowledge flow, innovation, and competitiveness; Ceipek et al. (2021) study the impact of family management on digital transformation, noting that family managed firms' conservatism, emotional attachment to existing assets, and rigid mindset can limit firms' investment in exploratory IoT innovations; Kortuem & Kawsar(2010) emphasize the importance of user innovation in the development of IoT, suggesting that user-led innovation can be fostered through the provision of a tool suite and an open-market mechanism and that IoT, like the smartphone app store like smartphone app stores to stimulate innovation. Paiola et al. (2022) analyze how SMEs can develop new IoT service-oriented business models within their existing business models, finding that this process is incremental and requires fine-tuning based on trial-and-error learning. Rossi et al. (2022) examine how public innovation intermediaries can adapt their business models to support the digital transformation of IoT , showing that these intermediaries transform from traditional technology upgraders and supporters to innovation system builders by reconfiguring their value propositions, activity organization, and target groups; Santoro et al. (2018) explor how knowledge management systems in the context of the IoT can enhance firms' innovation capabilities by facilitating open innovation and knowledge management capabilities.

Regarding IoT technology applications and challenges, Sollins(2019) explores the conflict between security and privacy requirements and innovation needs for IoT big data in his study, proposing a coordinated approach to decompose the design space from three dimensions: regulatory policy, economic business, and technological design; Mani & Chouk(2018) explore the factors that contribute to consumers' resistance to smart services and found that that complexity, security, health risks, and inconsistency with consumers' self-image trigger consumer resistance to IoT-based smart service innovations; Wu & Wang(2018) propose a distributed IoT security detection method based on game theory, which achieves information sharing through consensus protocols, analyzes the adversarial relationship between attackers and defenders, and improves system security; Zeng(2022) constructs an IoT energy efficiency optimization model based on game theory, studied the energy efficiency optimization problem of wireless sensor nodes, and proposed a non-correlated parallel learning algorithm to improve the overall energy efficiency of the system.

2. Smart Product Innovation

Some studies have focused on the design and innovation of smart products and their service systems, and explored how to realize value co-creation through system frameworks, data-driven, and user interactions, etc. Yin et al. (2020) explores the formation, definition, and characteristics of a sustainable smart product innovation ecosystem, and proposed that the system facilitates the co-creation and sharing of value through the integration of multiple stakeholders, which provides new perspectives and practical guidance for smart product innovation. a new perspective and practical guidance. Raff et al. (2020) clarify the core characteristics and conceptual framework of smart products through a systematic literature review and comprehensive analysis, and classified smart products into four layers: digital, connected, responsive, and intelligent, which pointed out the direction of enterprise innovation. Zheng et al. (2018) propose a data-driven smart product service system based on a data-driven design framework, which realizes the service innovation and value creation of intelligent products through the stages of platform development, data collection and analysis. Zhang et al. (2023) study the service innovation of the customer-product interaction life cycle in the intelligent product service system, constructed a three-dimensional service system, and proposed a personalized service recommendation method, which pushed forward the improvement of the theory of intelligent product service system. Bu et al. (2020) propose a hybrid intelligence approach to realize smart product service innovation by combining product user data and sensor data, and demonstrated the effectiveness of the approach in enhancing user engagement and real-time monitoring through case studies. Walk et al. (2023) investigate the application of deep learning in computer vision to enhance the sustainability of the manufacturing industry by detecting product wear state to support a sustainable smart product service system.

Other studies have deeply analyzed problems in complex systems based on game theory and proposed optimization methods to cope with uncertainty or improve efficiency. Xin et al. (2022) and others explore the service innovation of cloud manufacturing platforms from the perspective of multi-intelligence body game, constructed a tripartite evolutionary game model, and analyzed the impact of each subject's strategy on the service innovation of cloud manufacturing system. TM Rofin et al. (2022) propose an intelligent technology analysis method based on game theory for coping with supply chain disruptions during the New Crown Epidemic, which provides decision support for supply chain management. Liu et al. (2024) combine artificial intelligence with game theory to propose an innovative method to improve the efficiency of target tracking in sensor networks, which provides a new idea for the development of sensor network technology. Tsang et al. (2022) propose an intelligent product design framework that combines fuzzy association rule mining and genetic algorithm for the fuzzy front-end stage in new product development, providing a new approach for product innovation in manufacturing.

It is clear that most studies on innovation collaboration have paid less attention to how manufacturers and IoT platforms make decisions when collaborating on innovation at the same time. Our study aims to bridge this gap by focusing on the collaborative interactions between the two, thus providing some new insights for organizations.

3. Overconfidence

In the field of supply chain management, many scholars have thoroughly studied the multidimensional effects of overconfidence of manufacturers, retailers and other subjects on their decision-making behaviors and the overall efficiency of the supply chain. Li(2019) study explores the effects of overconfidence on manufacturers' and retailers' decisions and profits in a decentralized channel, and found that overconfidence reduces the channel's double marginal effect in some cases, and even makes the decentralized channel outperform the centralized channel. Zhou et al.(2022) et al. explores the impact of manufacturer overconfidence on green supply chain decision making and found that overconfidence leads to greener products and higher prices but lower profits, and investigated the role of revenue-sharing contracts in coordinating a green supply chain. Du et al.(2021) point out that in supply chain innovation scenarios, a manufacturer's overconfidence may have an impact on suppliers' investment in innovation and overall supply chain profitability, with the effect of the impact depending on the type of incentive contract used. Lu et al.(2023) investigate how retailers' overconfidence affects supply chain transparency, particularly in the context of manufacturers' potential access to the retail channel, and found that overconfident retailers are more motivated to improve supply chain transparency, while the opposite is true under the agency selling model. Xu et al.(2019) analyze pricing and ordering decisions between an overconfident retailer and a rational retailer in a duo-oligopoly supply chain, and find that overconfidence does not necessarily impair supply chain performance, and that the retailer's level of overconfidence affects its pricing and ordering decisions.

In the field of marketing and consumer behavior, Mahajan(1992) study shows that overconfident managers overestimate the accuracy of their predictions when making strategic marketing forecasts, but that overconfidence can be reduced and the accuracy of predictions can be improved by providing feedback, counterfactual reasoning, etc. Grubb(2009) points out that consumers' overconfidence leads them to underestimate their uncertainty about their demand for cell phone services, thus paying higher fees for initial consumption and facing high additional fees when they exceed their packages. Markovitch et al.(2015) explore how overconfidence affects managerial decision-making during the commercialization of start-ups and found that overconfidence is associated with overprediction of new product sales and that this effect is mediated. Razmdoost et al.(2015) examine the impact of overconfidence and lack of confidence on the dimensions of value perceived by consumers during the product or service use phase through two studies, which showed that both overconfidence and lack of confidence negatively affect different dimensions of consumer value.

This paper aims to further expand the boundaries of the research on manufacturer-platform collaboration in the smart product domain. By constructing a theoretical model, this paper analyzes manufacturers' decision-making behaviors and their impact on supply chain performance in usage-based contracts and revenue-sharing contracts under two scenarios: without overconfidence and with overconfidence. The study focuses on how manufacturers and platforms can adjust product pricing, innovation level, and contract terms to optimize the overall efficiency and profit distribution of the supply chain under the condition of non-perfect rationality. This study provides new perspectives for understanding the complex interactions between manufacturers and platforms in smart product supply chains, and offers theoretical support and practical guidance for actual business decisions.

Our study makes a significant contribution to the body of research on smart products, especially in the area of complementary innovations. Table 1 summarizes the comparison of our study with several key existing studies. Table 1 presents a comparative analysis of our study with several key existing studies, focusing on model construction in three key dimensions. It is evident from this comparison that there is still a small amount of literature encouraging manufacturer-platform collaboration in IoT environments. In addition, most scholars have yet to focus on whether manufacturers are fully rational or not and the impact that consumer type has on the supply chain.

Table 1 Comparison with existing model building studies

| Existing studies | Internet of Things Innovation | Smart Product Innovation | Overconfidence | Research methodology |
|---|---|---|---|---|
| Mani & Chouk(2018) | √ | | | empirical research |
| Yin et al. (2020) | | √ | | literature review |
| Xu et al.(2019) | | | √ | game model |
| (Du et al., 2021) | | | √ | game model |
| Haaker et al. (2021) | √ | | | case analysis |
| Zhou et al.(2022) | | | √ | game model |
| Zeng(2022) | √ | | | game model |
| Xin et al. (2022) | √ | √ | | game model |
| TM Rofin et al. (2022) | | √ | | game model |
| Lu et al.(2023) | | | √ | game model |
| Our paper | √ | √ | √ | game model |

Our paper is most closely related to the study by (Du et al., 2021), which examines the impact of overconfident downstream manufacturers on innovation and profitability of the entire supply chain, wholesale price contracts, and upstream suppliers under cost-sharing contracts for both them and innovative upstream suppliers, and explores what kind of contracts are taken to have more efficiency gains for the entire supply chain. And based on the existing literature, there is a distinct lack of research focusing on the IoT supply chain perspective. Our study focuses on the emerging area of IoT supply chain and fills the gap in existing research by linking IoT platforms and smart device manufacturers with the medium of collaborative innovation. Not only do we model innovations in manufacturer-owned smart products, but we also study platform investment strategies from the perspective of innovation co-creation and willingness to share consumer data triggered by information security. Specifically, our study focuses on the decision-making behavior of manufacturers and platforms in usage-based versus revenue-sharing contracts. In the absence of overconfidence, we analyze the profit functions of manufacturers and platforms, considering the impact of the proportion of non-privacy-sensitive consumers on the revenue and innovation investment of each party in the supply chain, and thus go beyond merely exploring the impact on innovation and profit in a situation where manufacturers are perfectly rational. In addition, our study further explores the impact of manufacturer overconfidence on supply chain decisions. Here again, the degree of consumer sensitivity to data privacy in the market is taken into account, as well as simultaneous innovation by manufacturers and platforms rather than just unilateral innovation by those who are upstream or on either side.

## 3.The model

As shown in Figure 1, consider that an upstream IoT platform provides IoT services to a downstream manufacturer, and the two parties work together to research and develop a new generation of smart products, and after the successful development, the manufacturer produces and sells the smart products to the final consumers at a price $p$. Meanwhile, there are two pricing models among platforms and manufacturers, one is a usage-based contract, where the manufacturer pays a fee $w$ to the supplier for technical services, and the other is a revenue-sharing contract, where the two share the proceeds from the sale of the product to the consumer based on a ratio of $r$ and $1-r$.

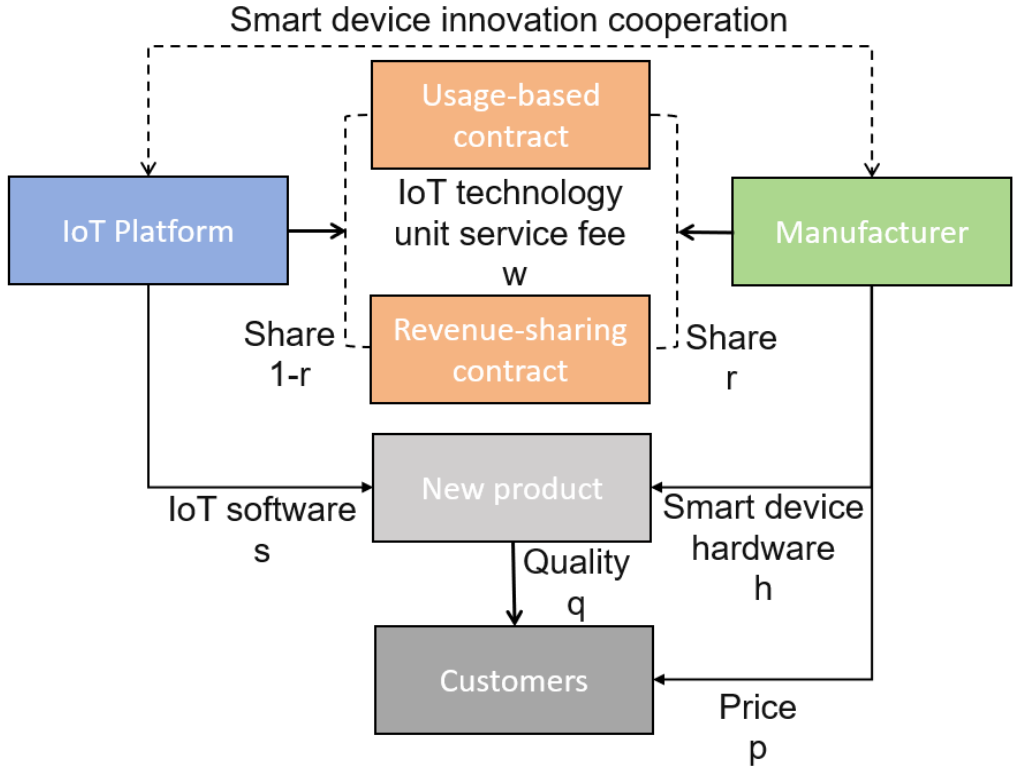

Figure 1 The model framework

## 3.1 Smart Products Innovation and Overconfidence

The platform mainly determines the level of innovation of its IoT service, denoted as s, by bearing a cost of $ks^2$ (Pei et al., 2023; Sun and Ji, 2022), and assumes that the non-innovation cost of the platform is zero. In the usage-based contract, we consider that the platform develops IoT services and sells them to the manufacturer at a fixed price $w$ per unit. In the revenue-sharing contract, we assume that the manufacturer can sell the smart product at a price $p$, and the platform's revenue share is $1-r$, so the platform receives a total revenue of $(1-r)p$. In working with the platform, we assume that the base functionality of the product supplied by the manufacturer is $q$. Also, for simplifying, we assume that the manufacturer's cost is 0. The manufacturer can decide on the level of hardware innovation for that product, denoted as $h$, and the cost of innovation borne by it is $kh^2$. Besides, we assume that the compatibility of hardware and software to be $\alpha$, then the total innovation level of the product is $Q = h + \alpha s$ (Chung et al., 2023; Noura et al., 2019). In the usage-based contract, since the manufacturer has already paid a fixed price to platform, the total revenue received by the manufacturer in this case is therefore the revenue received by the platform is $p$. In the revenue-sharing contract it is assumed that the share of revenue received by the manufacturer is $r$, and therefore the revenue received by the platform is $rp*D_t$.

In a real-world situation, we consider that manufacturers may be overconfident and set the magnitude of their level of overconfidence to $\varepsilon$. Overconfidence refers to the fact that individuals may overestimate their own knowledge and abilities, be more optimistic in their judgment of future situations, or underestimate others, believing that they are going to be better than others (Moore and Schatz, 2017), which can cause their expected benefits to deviate from what they are actually bringing in (Grubb, 2015). Although overconfidence may cause decision makers to overestimate future sales of new products(Markovitch et al., 2015b), in some cases supply chain transparency may increase due to retailers' overconfidence (Lu et al., 2023b). The manufacturer's overconfidence may be due to the fact that it acts as a seller in this segment, directly facing the sales market, and over-emphasizes its product advantages to compete with other sellers in the marketplace(Radzevick and Moore, 2011), thus increasing product pricing. It may also be due to insufficient or biased information available to the manufacturer when making pricing decisions, resulting in an overly optimistic estimate of market demand(Feiler and Tong, 2021), which may result in irrational pricing.

### 3.2 Smart device customers and Market Demand

Since consumers may have privacy sensitivities regarding the information shared for use in the IoT technology of the smart products, such consumers will not enjoy the value brought by the IoT technology. Based on this, we believe that there are two types of customers in the real situation, one is non-privacy-sensitive customers, and we set the proportion of this type of customers as $\lambda$, and the other is privacy-sensitive customers, which accounts for a proportion of the number of customers as $1-\lambda$. Customers s' sensitivity to product innovations is $\beta$, and due to the uncertainty in the level of innovations, product quality, and customers acceptance, we set this variable as a random variable with mean $\mu$ and variance $\sigma^2$ to describe the uncertainty of this variable in the range of $(0, \hat{\beta})$ with probability density function $f(\beta)$ and cumulative distribution function $F(\beta)$. From this we can derive the utility level $U_i = \gamma q + \beta(h + \alpha s) - p$ for non-privacy sensitive customers and $U_s = \gamma q + \beta h - p$ for privacy sensitive customers. the thresholds for the purchase of this smart product for the two types of customers are found to be $\gamma_i = \frac{p - \beta(h + \alpha s)}{q}$ and $\gamma_s = \frac{p - \beta h}{q}$, respectively, by making the utility level of the two types of customers equal to zero. Non-privacy-sensitive customers will choose to buy the product when their valuation is greater than $\gamma_i$ and privacy-sensitive customers will choose to buy the product when their valuation is greater than $\gamma_s$, therefore, $D_i = \left[1 - \frac{p-\beta(h+\alpha s)}{q}\right] \times \lambda$, and $D_s = \left(1 - \frac{p-\beta h}{q}\right) \times (1-\lambda)$, and we derive the total market demand for the smart product to be $D_t = \frac{-p+q+\beta(h+s\alpha\lambda)}{q}$.

### 3.4 Timeline for the Game

The game is divided into three phases, as shown in Figure 2. At the beginning of the game, the manufacturer and the platform first determine to use usage-based contracts and revenue-sharing contracts to accomplish transactional cooperation between them. The second stage is for the manufacturer and the platform to determine their own innovation level respectively, and then the manufacturer sells the smart product at price $p$. Finally the customers decides whether to buy the product or not.

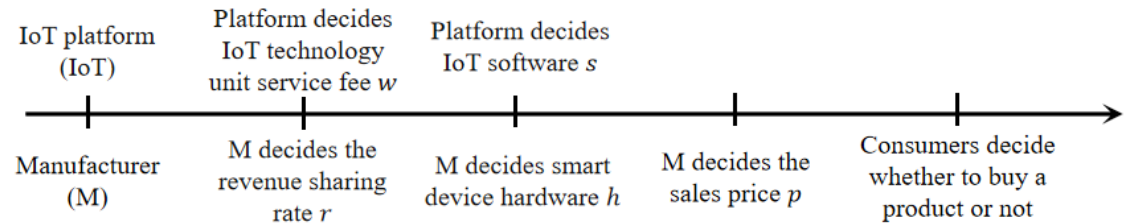

Figure 2 Timeline for the Game

In addition, we assume that the game participants are risk-neutral and make decisions based on their profit maximization. The table lists the key parameters and variables in our model.

Table 2 Summary of the Notation

| Notation | Definition |
|---|---|
| Indexes | |
| m | Index of manufacturers |
| p | Index of platform |
| sc | Index of supply chain |
| Parameters | |
| q | Base product quality |
| $c_m$ | Marginal cost of base product produced by the manufacturer |
| $c_s$ | Marginal cost of base product produced by the platform |
| $\gamma$ | Customers' marginal valuation of base quality |
| $\lambda$ | Proportion of privacy-indifferent customers |
| $\beta$ | Customer sensitivity to innovation |
| k | Demand spillover intensity |
| $\varepsilon$ | Manufacturer overconfidence levels |
| Decision variables | |
| p | Price of the product |

| h | Level of innovation of the manufacturer |
| r | Manufacturer's share in revenue sharing contract |
| w | Cost at which the manufacturer purchased the software |
| s | Level of innovation of the platform |
| Calculated values | |
| $D_s$ | Privacy-sensitive customer needs |
| $D_i$ | Privacy-indifferent customer needs |
| U | Customer utility |
| $\pi_m/\pi_s/\pi_{sc}$ | Profits of manufacturers/platform/supply chain |

## 4. Results and analysis

In this subsection, we analyze the decisions made by manufacturers and platforms in usage-based contracts versus revenue-sharing contracts, respectively, by considering the manufacturer's decisions in both the no-overconfidence and overconfidence scenarios, using backward induction.

4.1 No overconfidence

First, we analyze the profit function faced by the manufacturer and the platform respectively in the usage-based contract as

$$\max_{p,h} E[\pi_m^{un}(p,w,h,s)] = \int_0^{\widehat{\beta}} (p-w) D_t f(\beta) d\beta - kh^2 \quad (1)$$

$$\max_{w,s} E[\pi_p^{un}(p,w,h,s)] = \int_0^{\widehat{\beta}} w\, D_t f(\beta) d\beta - ks^2 + \theta s D_i \quad (2)$$

The corner symbol $un$ denotes the case where both the manufacturer and the platform are perfectly rational in the usage-based contract. Based on the previous analysis, we believe that non-privacy-sensitive customers are willing to share their information in using the software services in the smart product, and the platform can push other products and services in the platform to them based on the information obtained from the users, and thus can obtain additional profits from the customers. The benefits from this sharing of information should be proportional to the number of non-privacy-sensitive customers and the size of the software innovations included in the smart product. Thus, we have $\theta s D_i$, where $\theta$ denotes the marginal benefit that the platform derives from customer data sharing. Based on equations (1) and (2) we apply backward induction to find equilibrium solution $p^{un}$, $w^{un}$, $s^{un}$, $h^{un}$. The result is given in Lemma 1. To simplify the subsequent expressions, we make $A = \alpha\theta(\lambda-1)\lambda\mu$, $B = kq(8kq - \theta^2\lambda^2 + 2\alpha\theta(4-3\lambda)\lambda\mu + (2+\alpha^2\lambda^2)\mu^2)$.

Lemma 1: In the absence of overconfidence on the part of the manufacturer and when both parties opt for a usage-based contract, we can derive the manufacturer's equilibrium pricing, level of hardware innovation, and level of profitability for the smart product to be, respectively $p^{un} = \frac{q(6k^2q^2 - A\mu^2 - kq(\theta^2\lambda^2 + \mu^2 + \alpha\theta\lambda\mu - 5A))}{-2\mu^2 A + B}$, $h^{un} = \frac{q\mu(kq+A)}{-2\mu^2 A+B}$, $\Pi_m^{un} = \frac{kq^2(kq+A)^2(4kq-\mu^2)}{(-2\mu^2 A+B)^2}$, The platform's equilibrium revenue, software innovation level, and profit level for smart products are $w^{un} = \frac{q(4k^2q^2 - A\mu^2 - kq(\theta^2\lambda^2 + \mu^2 + \alpha\theta\lambda\mu - 3A))}{-2\mu^2 A + B}$, $s^{un} = \frac{kq^2\lambda(\theta+\alpha\mu)}{-2\mu^2 A+B}$, $\Pi_p^{un} = \frac{q\mu(kq+A)}{-2\mu^2 A+B}$.

In the revenue sharing contract, we have

$$\max_{p,h} E[\pi_m^{rn}(p,w,h,s)] = \int_0^{\hat{\beta}} r_n p\, D_t f(\beta) d\beta - kh^2 \qquad (3)$$

$$\max_{s} E[\pi_p^{rn}(p,w,h,s)] = \int_0^{\hat{\beta}} (1-r_n) p\, D_t f(\beta) d\beta - ks^2 + \theta s D_i \qquad (4)$$

The corner symbol $rn$ denotes the case where both the manufacturer and the platform are perfectly rational in the revenue-sharing contract. Based on equations (3) and (4) we apply backward induction to find the equilibrium solution $p^{rn}$, $s^{rn}$, $h^{rn}$, and the result is shown in Lemma 2. Again, to simplify the expression, we make $C = 16k^3q^3 + r_n^2\alpha\theta(1-\lambda)\lambda\mu^5 + kqr_n\mu^3(2\alpha\theta(4-3\lambda)\lambda + r_n\mu) - 4k^2q^2\mu(2r_n\mu + \alpha\lambda(2\theta(2-\lambda) + (1-r_n)\alpha\lambda\mu))$, $D = (4k^2q^2 + r_n\alpha\theta(1-\lambda)\lambda\mu^3 - kq\mu(\alpha\theta(4-3\lambda)\lambda + r_n\mu))$.

Lemma 2: In the absence of overconfidence on the part of the manufacturer and when both parties choose a revenue-sharing contract, we can derive the manufacturer's equilibrium pricing and level of hardware innovation for smart products to be, respectively $p^{rn} = \frac{2kq^2 D}{C}$, $h^{rn} = \frac{qr_n\mu D}{C}$, $\Pi_m^{rn} = \frac{kq^2 r_n(r_n\mu^2 - 4kq)D^2}{C^2}$. The level of software innovation and profitability of the platform for smart products are respectively $s^{rn} = \frac{kq^2\lambda(4kq(\theta+(1-rn)\alpha\mu) - r_n\theta\mu^2)}{C}$, $\Pi_p^{rn} = \frac{k^2q^3(4kq(1-r_n)+\theta\lambda(\theta\lambda+4(1-r_n)\alpha(\lambda-1)\mu))}{C}$.

Since the expression is too complicated, we assign certain parameters as follows, so that $\alpha = 1, q = 0.5, k = 0.5$, and we assume that $\theta \leq \frac{\mu}{2}$ since the benefit of information sharing for the platform is finite at the time.

**Proposition1.** In both contracts, when the manufacturer is perfectly rational, we have $\frac{\partial h}{\partial \lambda} > 0, \frac{\partial s}{\partial \lambda} > 0; \frac{\partial p}{\partial \lambda} > 0, \frac{\partial w}{\partial \lambda} > 0; \frac{\partial \pi_m}{\partial \lambda} > 0, \frac{\partial \pi_p}{\partial \lambda} > 0$.

Proposition 1 states that under both usage-based and revenue-sharing contract models, when the manufacturer is perfectly rational, the manufacturer's and the platform's profits, level of innovation, and product pricing all increase as the proportion of non-privacy-sensitive customers, $\lambda$, increases. This implies that when the proportion of non-privacy-sensitive consumers in the market rises, the revenue and innovation inputs of the supply chain parties increase in tandem, while product prices and software pricing costs rise due to higher demand or value recognition.

Proposition 1 suggests that when the proportion of non-privacy-sensitive customers increases, platforms are able to acquire more user data to optimize their services, which makes platforms more willing to invest in innovation in software technology, which allows consumers to derive higher utility from a smart product, which increases their demand for that product, and the value of the platform's innovations to the manufacturer increases significantly through this channel, and therefore platforms are more capable of bargaining for higher pricing of software products. Manufacturers may realize that as the proportion of non-privacy-sensitive customers increases, platforms can benefit from increased software innovation, and therefore manufacturers increase their hardware innovation in order to further increase market demand for smart products. At the same time, while manufacturers pay higher software innovation costs, they can set higher retail prices through higher levels of innovation and pass some of those costs on to consumers who value data services. As a result, platform and manufacturer costs can also increase with the number of sensitive customers.

For example, the practice of Amazon's Echo smart speaker and Alexa ecosystem perfectly confirms the above proposition. As the proportion of non-privacy-sensitive consumers continues to rise, Amazon continues to optimize its AI services (e.g., introducing generative AI to enable natural conversations) through the massive user data collected by the Alexa platform, which significantly improves the user experience and enables the platform to develop higher value returns for its software services. At the same time, manufacturers not only launched innovative hardware (e.g., Echo Show with screen, high-end audio devices), but also successfully passed on some of the costs by increasing retail prices, ultimately realizing the synchronous improvement of platforms and manufacturers in terms of profitability, innovation level and product pricing, which fully reflects the overall driving effect of the growth in the proportion of non-privacy-sensitive users on the value of the supply chain.

To facilitate the comparison of the size of the profits of the manufacturer and the platform under the two contracts, we let $\theta = 0.4$, $\mu = 1$ for taking values.

**Proposition2**. (1)For manufacturer, when $r_n < r_{n_1}$, we have $\pi_m^{un} > \pi_m^{rn}$; when $r_{n_1} < r_n < r_{n_2}$, $0 < \lambda < \lambda_1(r_n)$, we have $\pi_m^{un} > \pi_m^{rn}$, $\lambda_1(r_n) < \lambda < 1$, we have $\pi_m^{un} < \pi_m^{rn}$, when $r_{n_2} < r$, we have $\pi_m^{un} < \pi_m^{rn}$.

(2)For the platform, when $r_n < r_{n_3}$, $\pi_p^{rn} > \pi_p^{un}$; when $r_{n_3} < r_n < r_{n_4}$, $0 < \lambda < \lambda_2(r_n)$, we have $\pi_p^{un} > \pi_p^{rn}$, $\lambda_2(r_n) < \lambda < 1$, we have $\pi_p^{un} < \pi_p^{rn}$, when $r_{n_4} < r_n$, we have $\pi_p^{rn} < \pi_p^{un}$.

Proposition 2 states that the size $s$ of the profits obtained by the manufacturer and the platform in the two contract situations is affected by the share ratio $r_n$ and the proportion of non-privacy-sensitive customers $\lambda$. When the share ratio of the manufacturer or the platform is small, it is obvious that it is more favorable for the two to choose the contract based on the amount of usage. However, when the share ratio $r_n$ is in a certain interval, although it can be obtained according to Proposition 1 that the profit gained by the two increases with the proportion of non-privacy-sensitive customers, the different speeds of increase make the profit of the two contracts appear at an equilibrium point. Figure 3 illustrates how the manufacturer's profit varies with the increase in the proportion of non-privacy sensitive under both contracts when $r_n = 0.3$.

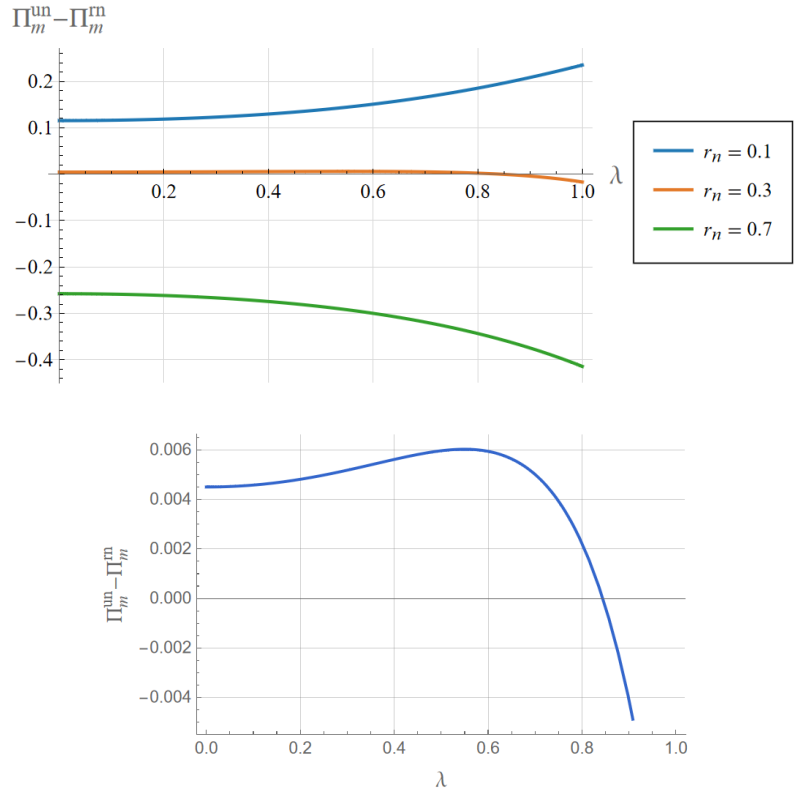

Figure 3 Manufacturer's profit

Proposition 2 shows that when $r_n < r_{n_1}$, the manufacturer's share of the revenue sharing contract is too low, which leads to low revenue for the manufacturer in this case, so the manufacturer chooses to buy software services at a fixed price based on the usage-based contract; when $r_{n_1} < r_n < r_{n_2}$, the improvement of the share of the revenue sharing contract makes the revenue sharing contract feasible. However, when the proportion of non-privacy-sensitive customers is small, the demand for software innovation is limited, so the manufacturer chooses to fix the innovation cost of software based on the usage contract to avoid the loss of revenue sharing, and at the same time improves the hardware innovation to increase the demand of privacy-sensitive customers to obtain higher profits; when the increase of non-privacy-sensitive customers can make the price rise by a large margin, then the choice of revenue sharing contract can incentivize the platform and the manufacturer to increase the price of software. When the increase in non-privacy-sensitive customers can increase prices by a large margin, the choice of a revenue-sharing contract can incentivize platforms and manufacturers to innovate in tandem, thereby effectively expanding consumer demand and increasing profits. For the platform, when $r_n < r_{n_3}$, the platform can get a larger share of the total revenue by virtue of the high share ratio of $1 - r_n$ in the revenue sharing contract, so the profit is higher. When $r_{n_3} < r_n < r_{n_4}$, the platform's share ratio decreases, at this time, if $\lambda < \lambda_2(r_n)$, a larger proportion of privacy-sensitive customers leads to limited demand and data sharing benefits from software innovation, the platform needs to bear the cost of software innovation may not be able to be covered by the benefits from the growth of demand, and therefore the choice of usage-based contract is more robust; if $\lambda_2(r_n) < \lambda$, the size of non-privacy sensitive customers further expands, so software innovation can bring more demand and data sharing benefits, and thus higher profits can be obtained by choosing the revenue sharing contract.

For example, the relationship between Apple's App Store and developers is a real-world example that illustrates proposition two. The platform (Apple) and manufacturers (developers) cooperate through revenue sharing contracts (30% or 15% commission share). When the share ratio is high (e.g., 30%), large developers believe that their own revenues have been excessively eroded, and tend to avoid the share, which triggers a contractual conflict; while when the share ratio is lowered to 15%, small and medium-sized developers, due to the significant growth in traffic and revenues brought about by the platform's innovations, are more willing to accept the share Contract. At the same time, if the scale of in-app paid users is small, developers are not motivated to choose the share contract; on the contrary, when the proportion of paid users expands, the share contract can effectively incentivize the platform to invest in innovation and bring higher demand, so that both sides can achieve profit equilibrium. This case clearly demonstrates how the split ratio and user structure dynamically affect the choice of supply chain contract and profit distribution.

### 4.2 Overconfidence

Based on the previous analysis, we believe that because the manufacturer plays the role of a seller in our study, due to the competitive needs or because of the insufficient information available when making decisions, there is an overestimation of the manufacturer's forecast of consumer demand, i.e., the manufacturer is overconfident in its level of innovation, and the level of overconfidence is $\varepsilon$. Since the marginal utility that the consumer obtains from the innovation of a smart product is $\beta$, when the manufacturer is overconfident, it believes that the utility that the consumer obtains for product innovation is $\beta_o = \beta + \varepsilon$. utility from smart product innovations is $\beta$, when the manufacturer is overconfident, where $\beta_0$ belongs to a random variable with mean $\mu + \varepsilon$ and variance $\sigma^2$. In this case, we still analyze the equilibrium decisions made by the manufacturer and the platform in both cases of usage-based contracts and revenue-sharing contracts.

Under a usage-based contract, the profit functions for the manufacturer and the platform are, respectively:

$$\max_{p,h} E[\pi_m^{uo}(p,w,h,s)] = \int_0^{\hat{\beta}} (p-w) D_t' f(\beta) d\beta - kh^2 \quad (5)$$

$$\max_{w,s} E[\pi_p^{uo}(p,w,h,s)] = \int_0^{\hat{\beta}} w D_t f(\beta) d\beta - ks^2 + \theta s D_i \quad (6)$$

The corner $uo$ denotes the presence of overconfidence on the part of the manufacturer in a usage-based contract. $D_t'$ denotes the market demand faced by the overconfident manufacturer, and since there is no overconfidence on the part of the platform, the market demand projected by the platform remains at $D_t$. Based on the previous analysis we consider $D_t = \frac{-p+q+\beta(h+s\alpha\lambda)}{q}$, and hence $D_t' = \frac{-p+q+(\beta+\varepsilon)(h+s\alpha\lambda)}{q}$. From equations (5) and (6) we can conclude that equilibrium solution $p^{uo}$, $w^{uo}$, $s^{uo}$, $h^{uo}$.

In the revenue sharing contract, we have

$$\max_{p,h} E[\pi_m^{ro}(p,w,h,s)] = \int_0^{\hat{\beta}} r_o p D_t' f(\beta) d\beta - kh^2 \quad (7)$$

$$\max_{s} E[\pi_p^{ro}(p,w,h,s)] = \int_0^{\hat{\beta}} (1-r_o) p D_t f(\beta) d\beta - ks^2 + \theta s D_i \quad (8)$$

The corner $ro$ denotes the presence of overconfidence of the manufacturer in the revenue sharing contract. According to equations (7) and (8) we can similarly find equilibrium solutions for $p^{ro}$, $s^{ro}$ and $h^{ro}$.

For ease of analysis, we similarly let $\alpha = 1, q = 0.5, k = 0.5$ in this case and let $\mu = 1$, thus $\varepsilon' = 1 + \varepsilon$ (for subsequent convenience of presentation, $\varepsilon'$ will be denoted by $\varepsilon$). Meanwhile, to ensure that the result has an equilibrium solution, the value of $\varepsilon$ ranges from $(1,2)$.

Proposition3. In the presence of manufacturer overconfidence in usage-based contracts, we have $\frac{\partial p^{uo}}{\partial \varepsilon} > 0, \frac{\partial w^{uo}}{\partial \varepsilon} < 0; \frac{\partial h^{uo}}{\partial \varepsilon} > 0, \frac{\partial s^{uo}}{\partial \varepsilon} < 0$.

**Proposition 3** indicates that the price and the level of hardware innovation determined by the manufacturer in usage-based contracts increase with its level of overconfidence, while the price of the software service and its level of innovation determined by the platform decreases with the level of manufacturer overcautiousness.

Proposition 3 reveals the impact of manufacturer overconfidence on its and the platform's decisions. Under usage-based contracts, when manufacturers have overconfidence, their product pricing monotonically decreases with increasing $\varepsilon$. This stems from manufacturers' overestimation of consumers' sensitivity to the value of innovation, believing that their innovations for smart products can effectively increase prices, leading them to adopt more aggressive pricing strategies. At the same time, manufacturers' hardware innovation level increases with ε, reflecting the positive incentive of overconfidence on their innovation investment. However, the platform's response strategy shows a reverse change: although the manufacturer's overconfidence can lead to an increase in hardware innovation of its products, since the platform is worried that the manufacturer's irrational pricing will lead to a significant decrease in its sales demand, in order to reduce the risk posed by its overconfidence, the platform's investment in software innovation decreases with $\varepsilon$, which reflects the platform's rational adjustment based on actual demand. Due to the decline in software innovation, the platform ensures cooperation with the manufacturer so that the software service fee it decides decreases with increasing ε, and at the same time, it adopts a compensatory mechanism to offset the lower demand that may be triggered by the manufacturer's overpricing. And these relationships hold in the full $\lambda \in (0,1)$ interval, suggesting that changes in the proportion of privacy-sensitive customers do not change the foundational effects of overconfidence on decision variables and have less impact on the making decisions made by manufacturers and platforms.

For example, the manufacturer (Nokia) is clearly overconfident in its brand influence and hardware innovations (e.g., Lumia's camera technology), which has led to aggressive hardware investment and high pricing in an attempt to compete head-to-head with iOS and Android. On the other hand, the platform side (Microsoft) was more rational, worrying that Nokia's irrational pricing would lead to a shrinking market demand, and instead reduced the pace of software innovation in Windows Phone, while maintaining cooperation and hedging its risks by providing Nokia with a compensation mechanism called "Platform Support Payment". In this case, consumers' sensitivity to privacy was not a central factor, and the end result demonstrates that manufacturers' overconfidence can lead to a departure from their own aggressive strategies and the platform's conservatism, which can lead to the failure of eco-cooperation.

In order to determine the impact of the manufacturer's overconfidence on the innovation decisions made by the manufacturer and the platform in the revenue-sharing contract, we further make $\theta = 0.4, \lambda = 0.5$. At this point, the value of ε is in the range of $(1, \varepsilon_1)$ if $0 < r_o < 1$ is satisfied.

**Proposition4**. In a revenue-sharing contract, when there is overconfidence in the manufacturer, we have $\frac{\partial p^{ro}}{\partial \varepsilon} > 0; \frac{\partial h^{uo}}{\partial \varepsilon} > 0; \ r_{o_1} < r_o < r_{o_2}$ when $1 < \varepsilon < \varepsilon(r_o), \frac{\partial s^{uo}}{\partial \varepsilon} > 0$, when $\varepsilon(r_o) < \varepsilon < \varepsilon_1, \frac{\partial s^{uo}}{\partial \varepsilon} < 0, r_{o_1} < r_o < 1$, and $\frac{\partial s^{uo}}{\partial \varepsilon} > 0$ when $r_{o_2} < r_o < 1$.

Under the revenue-sharing contract, if the manufacturer has overconfidence, the product pricing $p^{ro}$ and hardware innovation level $h^{uo}$ increase monotonically with $\varepsilon$. The direction of change of the platform software innovation level $s^{uo}$ is affected by its revenue-sharing ratio, and when $r_o$ is smaller than $r_{o_1}$, $s^{uo}$ shows a tendency to increase and then decrease with $\varepsilon$. When $r_o$ is larger than $r_{o_1}$, $s^{uo}$ improves with ε. uo increases with the increase of $\varepsilon$.

Proposition 4 suggests that when the platform share is higher (lower $r_o$), the manufacturer's product price and level of innovation increase with the level of overconfidence, regardless of the revenue it receives, possibly because when its revenue is lower, the manufacturer's overconfidence overestimates the innovation in software that is added by the platform that receives a higher revenue, and thus overestimates the demand in the market, and therefore it can subsequently increase its hardware innovation Therefore, it can then invest in hardware innovation and choose to set a higher price to increase its revenue, believing that this price increase will not cause much fluctuation in demand due to innovation. When an increase in their share of $r_o$ can more easily cover the cost of innovation, manufacturers may be more inclined to take the initiative to increase hardware innovation in order to set a higher price, overestimating market demand and price sensitivity in a moment of overconfidence.

When the platform's share is high and when the level of overconfidence is within a certain range, it is willing to follow the manufacturer's overconfidence to increase innovation because at this time the manufacturer does not deviate too much from rational pricing, and therefore it can obtain higher revenues to cover the cost of its innovations. However, when the overconfidence is higher, the platform, in order to circumvent the lower demand that may be caused by irrational behavior of the demand manufacturer, reduces the input of software innovations instead in order to reduce costs. However, as the manufacturer's share increases, the manufacturer's concern about covering innovation costs decreases and the manufacturer is more willing to invest in hardware innovation to raise prices. At this time, if the platform chooses to gradually reduce the level of innovation, it may have a greater impact on market demand, and in order to make pricing relatively rational, it chooses to follow the manufacturer's level of overconfidence to increase, and it can also increase the marginal return on information brought by non-privacy-sensitive customers. At the same time, the platform gradually reduces the level of software innovation while $r_o$ increases to ensure its own revenue, and the Figure 4 shows the changes in the level of software innovation with the manufacturer's overconfidence when $r_o$ is equal to 0.3, 0.5, and 0.7, respectively.

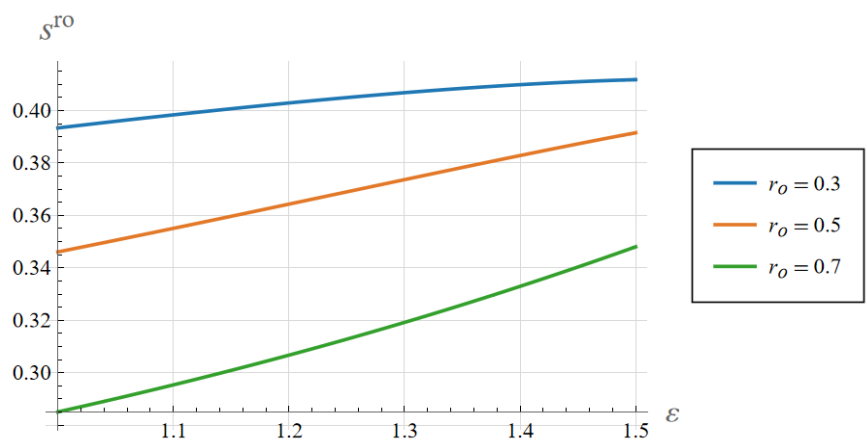

Figure 4 Level of innovation in software

For example, the interaction between Google's Android ecosystem and its hardware partners is a classic real-world example that illustrates proposition four. When the platform's share is high (e.g., when manufacturers such as Samsung need to share service revenues with Google), manufacturers' overconfidence drives them to adopt aggressive hardware innovations and high pricing strategies in order to boost their own revenues. At this point, if the manufacturer's level of overconfidence is moderate, the platform (Google) is willing to support software innovation (e.g., system adaptation), because success can bring ecological benefits; but if overconfidence is too high (e.g., out-of-control pricing), the platform becomes conservative and reduces the investment in dedicated innovation in order to avoid risk. However, when the manufacturer's share is high, as in the case of Google's own research, the platform, in order to avoid the impact of insufficient innovation on its own product demand, will instead simultaneously enhance software innovation (e.g., exclusive AI features), while gradually reducing the allocation of innovation resources to the general-purpose platform in order to maximize revenues. This case clearly demonstrates how the share ratio and the level of overconfidence work together to regulate the innovation game and risky decisions of platforms and manufacturers.

4.3 Comparison of overconfidence and no overconfidence

In the previous section, we analyze the decisions made by the manufacturer and the platform when the manufacturer has or does not have overconfidence, respectively, and next we will analyze the impact of the manufacturer's overconfidence on the decisions made by the manufacturer and the platform when they are facing a certain kind of contract. In the latter analysis, for the convenience of judging the size of the variables, we make $\alpha = 1, q = 0.5, k = 0.5, \theta = 0.4, \lambda = 0.5, \mu = 1$.

**Proposition 5**. In a usage-based contract, $\varepsilon$ takes the range $(1, \varepsilon_2)$ when $h^{uo} > h^{un}$, $s^{uo} < s^{un}$; $\pi_m^{uo} > \pi_m^{un}, \pi_p^{uo} > \pi_p^{un}$.

Proposition 5 shows that in usage-based contracts, when the manufacturer's level of overconfidence $\varepsilon$ is in the moderate interval $(1, \varepsilon_2)$, it invests in hardware innovation above the rational level, while the platform's software innovation is below the rational level. Nonetheless, both parties' profits are better than the benchmark when they are fully rational. This suggests that moderate overconfidence may enhance the overall supply chain returns through differentiated innovation strategies.

The proposition may be valid because, when $\varepsilon$ is in the interval $(1, \varepsilon_2)$, the manufacturer increases its investment in hardware innovation due to overestimation of consumers' sensitivity to innovation in an attempt to expand demand by enhancing product functionality, and under the fixed-fee model based on the usage-based contract, it can obtain all the benefits from hardware innovation, thus covering higher innovation costs and boosting its profitability; at the same time, while the platform avoids the risk of lower demand that may arise from manufacturer overpricing and chooses to reduce its costs by lowering the level of software innovation, and while the benefits of information sharing for non-privacy-sensitive customers may be lower as a result, the platform's revenue is guaranteed at a base level because of the fixed fee based on the volume of usage contract, and the expansion of demand due to hardware innovations more than offsets the shortfall in software innovations, and therefore the platform's revenue is are higher in the case of manufacturer overconfidence.

For example, DJI's model of cooperation with third-party software partners provides a realistic illustration of proposition five. As a manufacturer, DJI has demonstrated moderate overconfidence by continuously investing in hardware innovations (e.g., flight control systems, high-precision gimbals) that are higher than the industry's level of rationality, and has successfully explored the needs of emerging markets such as agricultural mapping and industrial inspection. On the other hand, software platforms have adopted a relatively conservative strategy, incrementally developing specialized applications based on actual needs (software innovation below the theoretical optimal level). In the framework of a usage-based contract (the software side pays a fixed SDK fee and subscribes on demand), DJI receives all the innovation revenues from the hardware sales, while the third party receives a stable profit from the market opportunities created by the hardware innovations. This case shows that the combination of moderate irrational innovation by the manufacturer and rational conservatism by the platform achieves overall supply chain gains through differentiated division of labor, confirming the conclusions of Proposition V.

**Proposition 6**. In a revenue sharing contract, $\varepsilon$ takes the range $(1, \varepsilon_1)$ and $r_n = r_o$. In this case $h^{ro} > h^{rn}$. When $r_{n_1} < r_n < r_{n_2}$ if $1 < \varepsilon < \varepsilon(r_n), s^{ro} > s^{rn}$, if $\varepsilon(r_n) < \varepsilon < \varepsilon_1, s^{ro} < s^{rn}$; when $r_{n_{2c}} < r_n < 1, s^{ro} > s^{rn}$; $\pi_m^{ro} > \pi_m^{rn}, \pi_p^{ro} > \pi_p^{rn}$.

Proposition 6 analyzes the impact of manufacturer's overconfidence on supply chain decisions and profits under revenue sharing contracts. The results show that when the manufacturer's overconfidence level is in the moderate range and the revenue-sharing ratio satisfies specific conditions, its hardware innovation level is always higher than the benchmark when it is perfectly rational, while the platform's software innovation level shows a nonlinear relationship with the share ratio: at higher platform share ratios, software innovation increases and then decreases, and when the platform's share ratio is lower, software innovation monotonically increases with $\varepsilon$. Notably, although the manufacturer's overconfidence may lead to irrational pricing, both parties' profits are better than the scenario when they are perfectly rational, suggesting that moderate overconfidence can achieve Pareto improvements in the supply chain under revenue-sharing contracts.

The conclusion of Proposition 6 reveals the positive effects of imperfectly rational behavior under a given contract. Manufacturers' overconfidence incentivizes consumers to increase their hardware innovation investment by overestimating their sensitivity to innovation, which enhances product differentiation and stimulates demand; when $r_o$ is small, platforms increase their innovation investment when the level of manufacturers' overconfidence is low, due to the fact that the platforms enjoy a higher revenue share and are able to gain more data revenue by cooperating with manufacturers' hardware innovation; however, as the level of overconfidence increases However, as the level of overconfidence increases, the manufacturer's pricing deviates from a reasonable range, and the platform reduces innovation to avoid the risk of demand shrinkage. When $r_o$ is large, as the manufacturer bears more innovation costs and revenue risks, its overconfident behavior always motivates the platform to maintain higher innovation inputs, because the platform's lower share makes it more dependent on the manufacturer's decisions, while the manufacturer's high share motivates it to expand the market demand through innovations, which creates a stable revenue expectation for the platform. And since the level of manufacturer overconfidence is in a reasonable range at this point $(\varepsilon_1 < \varepsilon_2)$, both returns are higher than those under no manufacturer overconfidence. The Figure 5 represents a comparison of platform software innovation under the profit-sharing contract when $r_o$ is equal to 0.1, 0.2, and 0.3, respectively.

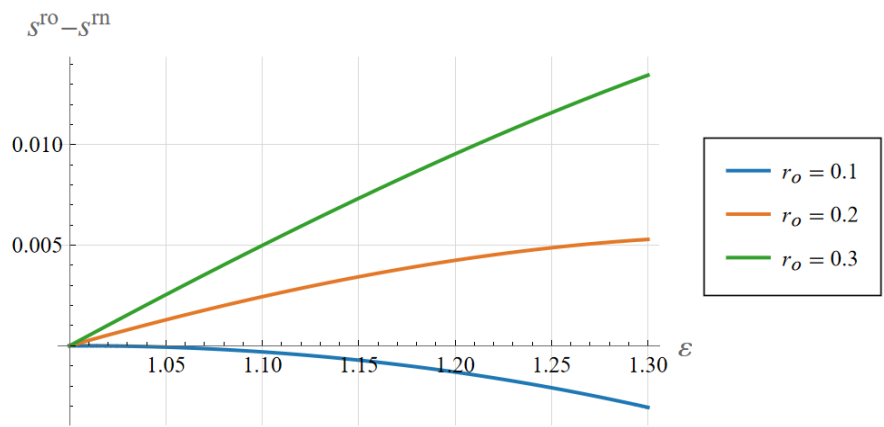

Figure 5 Level of innovation in software

For example, Huawei's HarmonyOS ecosystem and developer cooperation is a typical real-world example that illustrates proposition six. Under this revenue-sharing contract, Huawei, as a manufacturer, shows moderate overconfidence and continues to invest in hardware innovations (e.g., multi-device collaboration, Hongmeng distributed capabilities) above the industry's rational level in order to promote its "all-scene smart life" strategy. When the platform share is high, developers' software innovation increases and then decreases with Huawei's level of confidence: initially, they are willing to adapt to the Hongmeng system, but as Huawei's pricing strategy becomes more aggressive (e.g., high-end folded-screen positioning), some developers turn conservative to avoid demand risks. When Huawei increased the developer share through the "Star Program", developers' level of software innovation increased monotonically with Huawei's confidence level due to increased revenue security, and they were more active in developing native apps and deepening system integration. In the end, within the moderate confidence interval, both sides through hardware innovation to drive demand, software synergies to enhance the viscosity of the way, to achieve profits are higher than the level of the fully rational scenario.

## 5. Conclusions and implications

5.1 Conclusion

By constructing a game model that takes into account the fact that manufacturers and platforms innovate both hardware and software, that customers have different levels of privacy sensitivity to information, and that the marginal value of information brought by non-privacy-sensitive customers is introduced into the model, this study analyzes the impact of the manufacturer's self-confidence level on the supply chain decisions for smart products made by both it and the platform in different contractual modes. First, in both usage-based and revenue-sharing contracts, manufacturers' and platforms' profits, innovation levels, and product pricing are significantly affected by the proportion of non-privacy-sensitive customers. As the proportion of non-privacy-sensitive customers increases, the revenue and innovation investment of supply chain parties increase simultaneously, and product prices and software pricing costs rise. This finding suggests that changes in market demand structure have important guiding significance for the supply chain's innovation and pricing strategies, and managers should flexibly adjust the cooperation mode and innovation direction according to the target market characteristics in order to achieve the overall optimization of the supply chain.

Second, manufacturers' overconfidence has a complex double-edged sword effect on supply chain decisions. In usage-based contracts, moderate overconfidence can incentivize manufacturers to increase their hardware innovation investment, thus enhancing the overall revenue of the supply chain, but at the same time it may lead platforms to reduce the level of software innovation in order to avoid risks. In contrast, under revenue-sharing contracts, manufacturers' overconfidence positively incentivizes hardware innovation and pricing more significantly, while the platform's software innovation level is affected by the share ratio in a non-linear way. This result suggests that overconfidence can promote innovation and profit growth in the supply chain to some extent, but overconfidence may lead to irrational pricing and shrinking demand. Managers need to find a balance between incentivizing innovation and controlling risks, and rationally set the share ratio and innovation inputs to achieve robust supply chain development.

Finally, this study further reveals the positive role of imperfectly rational behavior in supply chains by comparing the decision differences when manufacturers are overconfident or not. In usage-based contracts, moderate overconfidence can enhance the overall return of the supply chain, while under revenue-sharing contracts, moderate overconfidence can even achieve Pareto improvement in the supply chain. This finding suggests that supply chain managers should not simply regard irrational behavior as a negative factor, but should guide partners' irrational behavior towards a change in the direction conducive to the development of the supply chain through flexible cooperation mechanisms and incentive strategies. At the same time, managers also need to pay close attention to changes in the market environment and optimize the information sharing mechanism to better coordinate innovation and pricing decisions in the supply chain, so as to enhance the competitiveness and sustainability of the supply chain.

5.2 Management Insights

Based on the above research findings, we draw the following management insights.

In supply chain management, choosing an appropriate cooperation model is the key to improving supply chain performance. By analyzing usage-based contracts and revenue-sharing contracts, this study finds that different contract models have a significant impact on manufacturers' and platforms' innovation inputs, profit levels, and pricing strategies. Managers should flexibly choose the cooperation model according to the structure of market demand and the characteristics of the partners. The usage-based contract is more suitable for stable and cost-sensitive demand, while the revenue-sharing contract can incentivize both parties to invest in innovation and improve the overall efficiency of the supply chain, especially when the proportion of non-privacy-sensitive customers is high. This revelation suggests that managers need to comprehensively consider the market environment and partner characteristics when choosing a collaboration model in order to optimize the supply chain and maximize its benefits.

Manufacturers' overconfidence has a complex effect in supply chain cooperation. Moderate overconfidence can incentivize manufacturers to increase innovation investment, which can enhance product differentiation advantages and market demand, but overconfidence may lead to irrational pricing and demand shrinkage, bringing risks to the platform. Managers need to channel partners' overconfident behavior through reasonable incentives and risk control measures. For example, in revenue-sharing contracts, manufacturers can be incentivized to remain rational in their innovation inputs and pricing strategies by adjusting their share ratios. At the same time, platforms also need to flexibly adjust their own innovation strategies according to manufacturers' overconfidence levels to achieve robust supply chain operations. This revelation suggests that managers should regard partners' irrational behavior as a potential innovation impetus and transform it into the competitiveness of the supply chain through effective management tools.

Changes in market demand structure have an important impact on the innovation strategies and pricing decisions of supply chains. This study finds that an increase in the proportion of non-privacy-sensitive customers can significantly increase the revenue and innovation investment of all parties in the supply chain, as well as prompt manufacturers and platforms to adjust their pricing strategies to meet market demand. Managers need to pay close attention to changes in the market environment, especially consumers' sensitivity to privacy and innovation, in order to adjust the innovation direction and cooperation mode of the supply chain in a timely manner. In addition, by optimizing the information sharing mechanism, managers can better coordinate innovation and pricing decisions in the supply chain to ensure that the supply chain's innovation strategy dynamically matches the market demand. This insight suggests that managers should focus on market orientation in supply chain management, and enhance the adaptability and competitiveness of the supply chain through flexible strategy adjustment and innovation management.

5.3 Future research directions

Despite the results of this study in exploring the impact of manufacturers' overconfidence on collaborative innovation in IoT supply chains, there are still some shortcomings. First, the study treats certain variables as fixed parameters, while in reality they may present dynamic characteristics over time and environmental changes, which may lead to an underestimation of the actual impact. Second, the study assumes that the proportion of non-privacy-sensitive customers is static, and does not fully consider the dynamic impact of privacy policies, technological development and other factors on customer behavior. Finally, in terms of contract design, the study does not deeply explore the dynamic adjustment and enforcement of contract terms, which may limit the applicability of the model in practical applications. Future research can construct a dynamic game model to further delve into the dynamics of manufacturers' overconfidence and its impact on long-term collaborative innovation in the supply chain, as well as introduce more behavioral economics theories to analyze the interactions of different behavioral traits and their impact on decision-making.

Conflicts of Interest

The authors declare no conflict of interest.

Data Availability Statement

The datasets used during the current study are available from the corresponding author on reasonable request

# Appendix

Lemma 1: under a usage-based contract, we can obtain the manufacturer's profit function as

$$\max_{p,h} E[\pi_m^{un}(p,w,h,s)] = \int_0^{\widehat{\beta}} (p-w)D_t f(\beta)d\beta - kh^2$$

The first order condition is $\frac{\partial \pi_m^{un}(p,w,h,s)}{\partial p} = 0$, $\frac{\partial \pi_m^{un}(p,w,h,s)}{\partial h} = 0$, it can be concluded that $p^{un} = \frac{-2(2+s^{un}\lambda\mu)+w^{un}(-2+\mu^2)}{-4+\mu^2}$, $h^{un} = \frac{\mu(-2+w^{un}-s^{un}\lambda\mu)}{-4+\mu^2}$. The second order condition is $H(p,h) = \begin{bmatrix} -1 & \frac{\mu}{2} \\ \frac{\mu}{2} & -1 \end{bmatrix}$. The determining factor is $1 - \frac{\mu^2}{4} < 0$

The profit function of the platform is

$$\max_{w,s} E[\pi_p^{un}(p,w,h,s)] = \int_0^{\widehat{\beta}} w\, D_t f(\beta)d\beta - ks^2 + \theta s D_i$$

The first order condition is $\frac{\partial \pi_p^{un}(p,w,h,s)}{\partial w} = 0$, $\frac{\partial \pi_p^{un}(p,w,h,s)}{\partial s} = 0$, it can be concluded that $w^{un} = \frac{q(4k^2q^2 - A\mu^2 - kq(\theta^2\lambda^2 + \mu^2 + \alpha\theta\lambda\mu - 3A))}{-2\mu^2 A + B}$, $s^{un} = \frac{kq^2\lambda(\theta+\alpha\mu)}{-2\mu^2 A + B}$. The second order condition is $H(p,h) = \begin{bmatrix} -\frac{2}{-4+\mu^2} & \frac{\lambda(\theta-\mu)}{-4+\mu^2} \\ \frac{\lambda(\theta-\mu)}{-4+\mu^2} & \frac{4-\mu^2+\theta\lambda\mu(-4+\mu^2-\lambda(-2+\mu^2))}{-4+\mu^2} \end{bmatrix}$. The determining factor is $-\frac{8-2\mu^2+2\theta\lambda\mu(-4+\mu^2-\lambda(-2+\mu^2))}{(-4+\mu^2)^2} < 0$. From this we can solve for $p^{un} = \frac{q(6k^2q^2 - A\mu^2 - kq(\theta^2\lambda^2 + \mu^2 + \alpha\theta\lambda\mu - 5A))}{-2\mu^2 A + B}$, $h^{un} = \frac{q\mu(kq+A)}{-2\mu^2 A + B}$. Since $\gamma_i$ and $\gamma_s$ are in the range (0,1), it follows that $\mu$ is in the range $\left(0, \frac{2}{\sqrt{3}}\right)$

The proof of Lemma II is the same as Lemma I. In order to satisfy the second order derivative of the two profit function of the main diagonal is less than 0, the determinant is greater than 0, and the range of values of $\gamma_i$ and $\gamma_s$ is (0,1), and the range of values of μ is $(0, Root[-28 + 24\#^2 + \#^4\&, 2, 0])$

Proposition 1 proves that, under both contracts, the manufacturer's innovation level $h$ and the product price have a partial derivative with respect to $\lambda$ $\frac{\partial h}{\partial \lambda} = -\frac{2\lambda\mu(\theta+\mu)^2(-2+\theta\lambda\mu)}{\left(-8+\theta^2\lambda^2+(2+\lambda^2)\mu^2+2\theta\lambda\mu(4-\mu^2+\lambda(-3+\mu^2))\right)^2}$, $\frac{\partial p}{\partial \lambda} = \frac{2\lambda(-2+\theta\lambda\mu)(4\theta\mu-\mu^2(-6+\mu^2)+\theta^2(-2+\mu^2))}{(-8+\theta^2\lambda^2+(2+\lambda^2)\mu^2+2\theta\lambda\mu(4-\mu^2+\lambda(-3+\mu^2)))^2} > 0$;

The platform innovation level s and pricing w pairwise $\lambda$ partial derivatives are $\frac{\partial w}{\partial \lambda} =$

$$-\frac{2\lambda(-2+\theta\lambda\mu)(\theta^2-\mu^2)(-4+\mu^2)}{\left(-8+\theta^2\lambda^2+(2+\lambda^2)\mu^2+2\theta\lambda\mu(4-\mu^2+\lambda(-3+\mu^2))\right)^2}>0, \frac{\partial s}{\partial \lambda}=$$

$$-\frac{2(\theta+\mu)\left(8+\theta^2\lambda^2+(-2+\lambda^2)\mu^2+2\theta\lambda^2\mu(-3+\mu^2)\right)}{\left(-8+\theta^2\lambda^2+(2+\lambda^2)\mu^2+2\theta\lambda\mu(4-\mu^2+\lambda(-3+\mu^2))\right)^2}>0;$$ The partial derivatives of both profits with respect to $\lambda$ are $\frac{\partial \pi_m}{\partial \lambda}=-\frac{4\lambda(\theta+\mu)^2(-2+\theta\lambda\mu)(1+\theta(-1+\lambda)\lambda\mu)(-4+\mu^2)}{\left(-8+\theta^2\lambda^2+(2+\lambda^2)\mu^2+2\theta\lambda\mu(4-\mu^2+\lambda(-3+\mu^2))\right)^3}>0, \frac{\partial \pi_p}{\partial \lambda}=$

$$-\frac{2\lambda(\theta+\mu)^2(-2+\theta\lambda\mu)}{\left(-8+\theta^2\lambda^2+(2+\lambda^2)\mu^2+2\theta\lambda\mu(4-\mu^2+\lambda(-3+\mu^2))\right)^2}>0.$$

Proof of Proposition 2: When values of $\theta$ and $\mu$ are taken, the manufacturer's profit $\pi_m^{un}-\pi_m^{rn}$ under the two contracts changes positively or negatively as $\lambda$ and $r_n$ change. When $r_n<r_{n_1}$, $\pi_m^{un}-\pi_m^{rn}$ is always greater than 0, for example when $r_n=0.2$, $\pi_m^{un}>\pi_m^{rn}$; when $r_n$ is in the range of $(r_{n_1},r_{n_2})$ and $0<\lambda<\lambda_1(r_n)$, there is $\pi_m^{un}>\pi_m^{rn}$; when $\lambda_1(r_n)<\lambda<1$, there is $\pi_m^{un}<\pi_m^{rn}$, for example, when $r_n=0.3$, $\lambda\in(0,Root[-791453125+1266325000\#-2159597500\#\^{}2+1882375000\#\backslash\^{}3+602171925\#\^{}4-945883300\#\^{}5+836904192\#\^{}6-251924024\#\backslash\^{}7+11038532\#\^{}8\&,2,0])$, $\pi_m^{un}>\pi_m^{rn}$; when $r_{n_2}<r_n$, there is $\pi_m^{un}<\pi_m^{rn}$.

Proof of Proposition 3: Based on the profit function of the two, applying backward induction can be solved for

In usage-based contracts

$$p^{uo}=\frac{-4(1+\varepsilon)(-6+\varepsilon^2)+4(1+\varepsilon)(-6+\varepsilon^2)\theta\lambda+2(4(-1+\varepsilon)-2(-3+\varepsilon)(1+\varepsilon)^2\theta+(-2+\varepsilon)(1+\varepsilon)^2\theta^2)\lambda^2}{(-2+\varepsilon)(-4(1+\varepsilon)(2+\varepsilon)+4(1+\varepsilon)(2+\varepsilon)\theta\lambda+(4+(1+\varepsilon)^2(-4+\theta)\theta)\lambda^2)},$$

$$h^{uo}=\frac{2\varepsilon(2(1+\varepsilon)-2(1+\varepsilon)\theta\lambda+(-2+\theta+\varepsilon(2+(2+\varepsilon)\theta))\lambda^2)}{(-2+\varepsilon)(-4(1+\varepsilon)(2+\varepsilon)+4(1+\varepsilon)(2+\varepsilon)\theta\lambda+(4+(1+\varepsilon)^2(-4+\theta)\theta)\lambda^2)},$$

$$w^{uo}=\frac{2(1+\varepsilon)(-2(2+\varepsilon)+2(2+\varepsilon)\theta\lambda+(1+\varepsilon)(-2+\theta)\theta\lambda^2)}{-4(1+\varepsilon)(2+\varepsilon)+4(1+\varepsilon)(2+\varepsilon)\theta\lambda+(4+(1+\varepsilon)^2(-4+\theta)\theta)\lambda^2},$$

$$s^{uo}=-\frac{2(1+\varepsilon)(2+\theta+\varepsilon\theta)\lambda}{-4(1+\varepsilon)(2+\varepsilon)+4(1+\varepsilon)(2+\varepsilon)\theta\lambda+(4+(1+\varepsilon)^2(-4+\theta)\theta)\lambda^2};$$

$$\pi_m^{uo}=-\frac{2(2+\varepsilon)\left(2(1+\varepsilon)-2(1+\varepsilon)\theta\lambda+\left(-2+\theta+\varepsilon(2+(2+\varepsilon)\theta)\right)\lambda^2\right)^2}{(-2+\varepsilon)(-4(1+\varepsilon)(2+\varepsilon)+4(1+\varepsilon)(2+\varepsilon)\theta\lambda+(4+(1+\varepsilon)^2(-4+\theta)\theta)\lambda^2)^2},$$

$$\pi_p^{uo}=-\frac{2(1+\varepsilon)^2(1+\theta(-1+\lambda)\lambda)}{-4(1+\varepsilon)(2+\varepsilon)+4(1+\varepsilon)(2+\varepsilon)\theta\lambda+(4+(1+\varepsilon)^2(-4+\theta)\theta)\lambda^2}.$$

Also, to ensure that $\gamma_i$ and $\gamma_s$ take values in the range $(0,1)$, $\varepsilon \in (1,Root[-12+12\theta\lambda+4\lambda^2-6\theta\lambda^2+2\theta^2\lambda^2+(-12+4\lambda+14\theta\lambda-4\lambda^2-10\theta\lambda^2+3\theta^2\lambda^2)\#1+(4+2\lambda-\theta\lambda-2\lambda^2-\theta\lambda^2)\#1^2+(4-2\lambda-4\theta\lambda+2\lambda^2+4\theta\lambda^2-\theta^2\lambda^2)\#1^3+(-\theta\lambda+\theta\lambda^2)\#1^4\&,3])$

In revenue-sharing contracts,

$$p^{ro}=\frac{16-4r_o\varepsilon^2+4(-4+r_o\varepsilon^2)\theta\lambda-2\varepsilon(2(-1+r_o)(-1+\varepsilon)+(-6+r_o\varepsilon(1+\varepsilon))\theta)\lambda^2}{(-4+r_o\varepsilon^2)^2-(-4+r_o\varepsilon^2)^2\theta\lambda+\varepsilon(-4(-1+r_o)(-2+\varepsilon)+(-2+r_o\varepsilon)(-4+r_o\varepsilon^2)\theta)\lambda^2},$$

$$h^{ro} = \frac{r_o\varepsilon(8-8\theta\lambda+\varepsilon(-2r_o\varepsilon+2r_o\varepsilon\theta\lambda-(2(-1+r_o)(-1+\varepsilon)+(-6+r_o\varepsilon(1+\varepsilon))\theta)\lambda^2))}{(-4+r_o\varepsilon^2)^2-(-4+r_o\varepsilon^2)^2\theta\lambda+\varepsilon(-4(-1+r_o)(-2+\varepsilon)+(-2+r_o\varepsilon)(-4+r_o\varepsilon^2)\theta)\lambda^2},$$

$$s^{ro} = \frac{(2(-1+r_o)(-4+r_o(-1+\varepsilon)\varepsilon^2)+(-2+r_o(-1+\varepsilon)\varepsilon)(-4+r_o\varepsilon^2)\theta)\lambda}{(-4+r_o\varepsilon^2)^2-(-4+r_o\varepsilon^2)^2\theta\lambda+\varepsilon(-4(-1+r_o)(-2+\varepsilon)+(-2+r_o\varepsilon)(-4+r_o\varepsilon^2)\theta)\lambda^2};$$

$$\pi_m^{ro} = -\frac{(r_o(-4+r_o\varepsilon^2)(-8+8\theta\lambda+\varepsilon(2r_o\varepsilon-2r_o\varepsilon\theta\lambda+(2(-1+r_o)(-1+\varepsilon)+(-6+r_o\varepsilon(1+\varepsilon))\theta)\lambda^2))^2}{2((-4+r_o\varepsilon^2)^2-(-4+r_o\varepsilon^2)^2\theta\lambda+\varepsilon(-4(-1+r_o)(-2+\varepsilon)+(-2+r_o\varepsilon)(-4+r_o\varepsilon^2)\theta)\lambda^2)^2},$$

$$\pi_p^{ro} =$$

$$\frac{8(-1+r_o)(-2+r_o(-1+\varepsilon)\varepsilon)-8(-1+r_o)(-2+r_o(-1+\varepsilon)\varepsilon)\theta\lambda+(4(-1+r_o)^2(-1+\varepsilon)^2+4(-1+r_o)(1+\varepsilon)(-2+r_o(-1+\varepsilon)\varepsilon)\theta+(-2+r_o(-1+\varepsilon)\varepsilon)^2\theta^2)\lambda^2}{2((-4+r_o\varepsilon^2)^2-(-4+r_o\varepsilon^2)^2\theta\lambda+\varepsilon(-4(-1+r_o)(-2+\varepsilon)+(-2+r_o\varepsilon)(-4+r_o\varepsilon^2)\theta)\lambda^2)}$$

. To ensure that $\gamma_i$ and $\gamma_s$ take values in the range $(0,1)$, when $\theta = 0.4, \lambda = 0.5$, $\varepsilon \in (1, Root[64 + 30\# - 38\#^2 - 17\#^3 + \#^4 \&, 3, 0])$

Proof of Proposition 4: $\varepsilon_1 = Root[64 + 30\# - 38\#^2 - 17\#^3 + \#^4 \&, 3, 0]$. ( Proposition 3 seeks an equilibrium solution to the revenue sharing contract obtained as $\varepsilon_1$)

When $\lambda = \frac{1}{2}, \theta = \frac{2}{5}, 0 < r_o < 1$,

$$\frac{\partial p^{ro}}{\partial \varepsilon}$$

$$= \frac{\begin{aligned}&2(ro^3\varepsilon^4(1+\theta(-1+\lambda)\lambda)(2\varepsilon\theta(-2+\lambda)\lambda-6\lambda^2+\varepsilon^2\theta\lambda^2+4\varepsilon(1+\lambda^2))-8\lambda^2(-4+4\theta^2\lambda+\varepsilon^2\lambda^2\\&+\theta(-4+4\lambda+\varepsilon^2\lambda^2))+8ro(\varepsilon^3\theta(1+\theta)\lambda^4+4\lambda^2(-1+\theta\lambda)+\varepsilon(8+4\theta(-4+\lambda)\lambda-4\theta^2(-2+\lambda)\lambda^2)\\&+\varepsilon^2\lambda^2(2\theta^2(-2+\lambda)\lambda+2\lambda^2+\theta(4+\lambda^2)))-2ro^2\varepsilon^2(4\lambda^4+2\varepsilon^3\lambda^2(1+\theta(-1+\lambda)\lambda)+\varepsilon^2\lambda^2(-3+\theta^2\lambda(-7+8\lambda)\\&+\theta(7+3\lambda-3\lambda^2))-4\varepsilon(-4+4\theta^2(-1+\lambda)\lambda^2-\theta\lambda(-8+4\lambda+\lambda^3))))\end{aligned}}{(ro^2\varepsilon^4(1+\theta(-1+\lambda)\lambda)+4(4-2\varepsilon\lambda^2+\varepsilon^2\lambda^2+2\theta\lambda(-2+\varepsilon\lambda))-2ro\varepsilon(-4\lambda^2+\varepsilon^2\theta\lambda^2+2\varepsilon(2+\theta(-2+\lambda)\lambda+\lambda^2)))^2}$$

$$\frac{\partial h^{ro}}{\partial \varepsilon}$$

$$= \frac{\begin{aligned}&ro\\&(((-4+ro\varepsilon^2)^2-(-4+ro\varepsilon^2)^2\theta\lambda+\varepsilon(-4(-1+ro)(-2+\varepsilon)+(-2+ro\varepsilon)(-4+ro\varepsilon^2)\theta)\lambda^2)\\&(8-8\theta\lambda+\varepsilon(-2ro\varepsilon+2ro\varepsilon\theta\lambda-(2(-1+ro)(-1+\varepsilon)+(-6+ro\varepsilon(1+\varepsilon))\theta)\lambda^2))+\varepsilon((-4+ro\varepsilon^2)^2\\&-(-4+ro\varepsilon^2)^2\theta\lambda+\varepsilon(-4(-1+ro)(-2+\varepsilon)+(-2+ro\varepsilon)(-4+ro\varepsilon^2)\theta)\lambda^2)(2(-1+2\varepsilon+3\theta)\lambda^2\\&-ro(2\varepsilon\theta(-2+\lambda)\lambda-2\lambda^2+3\varepsilon^2\theta\lambda^2+4\varepsilon(1+\lambda^2)))-\varepsilon(8-8\theta\lambda+\varepsilon(-2ro\varepsilon+2ro\varepsilon\theta\lambda-(2(-1+ro)(-1+\varepsilon)\\&+(-6+ro\varepsilon(1+\varepsilon))\theta)\lambda^2))(8(-1+\varepsilon+\theta)\lambda^2+4ro^2\varepsilon^3(1+\theta(-1+\lambda)\lambda)-2ro(-4\lambda^2+3\varepsilon^2\theta\lambda^2\\&+4\varepsilon(2+\theta(-2+\lambda)\lambda+\lambda^2))))\end{aligned}}{((-4+ro\varepsilon^2)^2-(-4+ro\varepsilon^2)^2\theta\lambda+\varepsilon(-4(-1+ro)(-2+\varepsilon)+(-2+ro\varepsilon)(-4+ro\varepsilon^2)\theta)\lambda^2)^2}$$

When $r_{o_1} < r_o < r_{o_2}$, when $1 < \varepsilon < \varepsilon(r_o), \frac{\partial s^{ro}}{\partial \varepsilon} > 0$, when $\varepsilon(r_o) < \varepsilon < \varepsilon_1, \frac{\partial s^{ro}}{\partial \varepsilon} < 0$, for example when $r_o = 0.15$, $1 < \varepsilon < \varepsilon(r_o), 1 < \varepsilon < Root[24144000 + 1424000\# - 18204600\#^{\wedge}2 + 280800\#^{\wedge}3 + 5700\#^{\wedge}4 + 2385\#^{\wedge}5 + 10152\#^{\wedge}6\&, 3, 0], \frac{\partial s^{ro}}{\partial \varepsilon} > 0$;

When $Root[24144000 + 1424000\# - 18204600\#^2 + 280800\#^3 + 5700\#^{\wedge}4 + 2385\#^5 + 10152\#^6\&, 3, 0] < \varepsilon < \varepsilon_1, \frac{\partial s^{ro}}{\partial \varepsilon} < 0$.

When $r_{o_2} < r_o < 1$, $\frac{\partial s^{ro}}{\partial \varepsilon} > 0$.

Proof of Proposition 5: the range of values of $\varepsilon$ is $(1, \varepsilon_2)$. $\varepsilon_2 = \text{Root}[-12 + 12\theta\lambda + 4\lambda^2 - 6\theta\lambda^2 + 2\theta^2\lambda^2 + (-12 + 4\lambda + 14\theta\lambda - 4\lambda^2 - 10\theta\lambda^2 + 3\theta^2\lambda^2)\#1 + (4 + 2\lambda - \theta\lambda - 2\lambda^2 - \theta\lambda^2)\#1^2 + (4 - 2\lambda - 4\theta\lambda + 2\lambda^2 + 4\theta\lambda^2 - \theta^2\lambda^2)\#1^3 + (-\theta\lambda + \theta\lambda^2)\#1^4 \&, 3]$. (Proposition III solves for $\varepsilon_2$ obtained based on the equilibrium solution of the usage contract)

Under these conditions $h^{uo} > h^{un}$, $s^{uo} < s^{un}$; $\pi_m^{uo} > \pi_m^{un}$, $\pi_p^{uo} > \pi_p^{un}$.

Proof of Proposition 6: The value of $\varepsilon$ ranges from $(1, \varepsilon_1)$ and $r_n = r_o$. In this case $h^{ro} > h^{rn}$. When $r_{n_1} < r_n < r_{n_2}$ if $1 < \varepsilon < \varepsilon(r_n), s^{ro} > s^{rn}$, if $\varepsilon(r_n) < \varepsilon < \varepsilon_1, s^{ro} < s^{rn}$; when $r_{n_{2c}} < r_n < 1$, $s^{ro} > s^{rn}$; $\pi_m^{ro} > \pi_m^{rn}, \pi_p^{ro} > \pi_p^{rn}$. For example when $r_n = 0.125$, $1 < \varepsilon < \text{Root}[-120928 + 54144\# + 36349\#^2 + 993\#^3\&, 3, 0], s^{ro} > s^{rn}$ ; when $\text{Root}[-120928 + 54144\# + 36349\#^2 + 993\#^3\&, 3, 0] < \varepsilon < \varepsilon_1, s^{ro} < s^{rn}$, $\pi_m^{ro} > \pi_m^{rn}, \pi_p^{ro} > \pi_p^{rn}$